# Neutron scattering by Dirac multipoles

S. W. Lovesey[1,2] and D. D. Khalyavin[1]

[1]ISIS Facility, STFC Oxfordshire OX11 0QX, UK

[2]Diamond Light Source Ltd, Oxfordshire OX11 0DE, UK

**Abstract** Scattering by magnetic charge formed by Dirac multipoles that are magnetic and polar is examined in the context of materials with properties that challenge conventional concepts. An order parameter composed of Dirac quadrupoles has been revealed in the pseudo-gap phase of ceramic, high-$T_c$ superconductors on the basis of Kerr effect and magnetic neutron Bragg diffraction measurements. Construction of Dirac quadrupoles that emerge from centrosymmetric sites used by Cu ions in the ceramic superconductor Hg1201 is illustrated, together with selection rules for excitations that will feature in neutron inelastic scattering, and RIXS experiments. We report magnetic scattering amplitudes for diffraction by polar multipoles that have universal value, because they are not specific to ceramic superconductors. To illustrate this attribute, we consider neutron Bragg diffraction from a magnetically ordered iridate ($Sr_2IrO_4$) and discuss shortcomings in published interpretations of diffraction data.

**1. Introduction**

Long-range magnetic order in a material is conventionally characterized by a motif of magnetic dipole moments [1]. The dipoles in question are expectation values of **μ** = (2**S** + **L**) where **S** and **L** are spin and orbital operators, respectively. In the event that dipole moments are zero the motif is often labelled a hidden magnetic-order, or perhaps symmetry protected, simply because it is not apparent in observations conducted in the laboratory. While dipole moments might be forbidden by symmetry, the same ruling need not apply to expectation values of magnetic tensors of higher rank, e.g., magnetic quadrupoles (rank 2) and magnetic octupoles (rank 3) might be different from zero. Fortunately, higher-rank multipoles are not beyond observation, because they deflect x-rays and neutrons [2, 3, 4].

Legislation as to the existence and properties of multipoles in a crystal is handed down by one, or more, of the 122 magnetic point-groups that delineate symmetry embedded in the environment of a site used by a magnetic ion. Magnetic multipoles are time-odd by definition. Multipoles mentioned in the previous paragraph, which constitute conventional magnetism, are also parity-even (axial). Bulk properties of a magnetic crystal, e.g., the Kerr effect, are prescribed by the magnetic crystal-class that is formed by the union of site symmetry (a point group) and translation symmetry in the motif. The Kerr effect can occur without a ferromagnetic motif of magnetic dipoles **μ** = (2**S** + **L**) that creates magnetic Bragg spots indexed on the chemical structure.

A second type of long-range magnetic order is composed of Dirac multipoles. These multipoles are products of time-odd electronic operators, **S** or **L**, and the time-even electric dipole operator, **n**, with products time-odd and parity-odd (magnetic and polar). [5, 6, 7] (Dirac multipole is an accurate and appropriate neologism present in literature past. Magneto-electric and magnetic charge are synonymous labels.) Recall that, **S** and **L** are pseudovectors – axial vectors do not change sign upon inversion of spatial coordinates – while **n** is a true polar vector. In this second scenario for magnetism, expectation values of anapoles (**S x n**) or **Ω** = (**L x n** − **n x L**) may replace expectation values of **µ** to define a magnetic motif. Long-range order contains both axial and polar multipoles, in the general case. However, axial magnetism can be entirely forbidden, in which case the magnetic state of the material is magnetic charge due solely to Dirac multipoles.

A state of pure magnetic charge in a material is the assured outcome when magnetic ions occupy sites that possess inversions of space and time as conjugate symmetry operations. In such cases, anti-inversion $\bar{1}'$ is an element of site symmetry and axial multipoles of any rank are forbidden. Of the 122 magnetic point-groups 21 contain anti-inversion $\bar{1}'$ and conventional, axial magnetism is absent.

Magnetic charge, epitomized by a magnetic monopole, is notable by its absence in Maxwell's equations that unite electricity and magnetism. Artificially inserted in the equations, with symmetries of the electric and the magnetic field unchanged, magnetic charge is both time-odd and parity-odd like Dirac's magnetic monopole. A Dirac monopole (**S • n**) can contribute in x-rays scattering [8], but it is forbidden by symmetry from contributing to neutron scattering [4]. A monopole using **L** does not exist, because **L** and **n** are orthogonal operators and (**L • n**) = 0.

A ferro-type motif of Dirac quadrupoles has recently been uncovered in the pseudo-gap phase of the cuprate Hg1201 by magnetic neutron diffraction [7, 9]. Analysis of similar neutron diffraction experiments on YBCO reveal the possibility of a combination of axial and polar magnetism that likewise does not break translation symmetry, although available neutron diffraction data, indeed, are fully explained by a ferro-type motif of pure polar magnetism [6, 10]. Previously, the most prominent feature of the normal state of cuprate superconductors is a loss with decreasing temperature of states at the Fermi energy. Given the large number of site symmetries that forbid axial magnetism and allow polar magnetism – 21 out of a total of 122 point groups – it seems unlikely that the pseudo-gap phase of Hg1201 is an isolated example of ordered magnetic charge due to Dirac multipoles. There are possibly a larger number of materials that support mixed axial and polar magnetism, which exists in YBCO and has been proposed as a candidate for hidden magnetic-order in orthorhombic FeSe [11].

The Kerr effect is allowed for both high-$T_c$ materials we mention, even though ferromagnetism is forbidden in the designated magnetic crystal-classes. Orenstein proposed that the Kerr effect and neutron Bragg diffraction data for the pseudo-gap phase could be reconciled with an antiferromagnetic order of local, canted magnetic moments of unknown origin [12]. A few years later, it was demonstrated theoretically that neutrons are scattered by Dirac multipoles [4], in addition to conventional axial multipoles. Shortly thereafter, successful analyses of enigmatic neutron diffraction data for Hg1201 and YBCO in terms of a ferro-type order of polar quadrupoles were published [6, 7]. Analyses in question are simple and straightforward, with use of standard methods of magnetic crystallography, once Dirac multipoles are admitted to the amplitude for magnetic neutron diffraction. Notably, reconciliation of available data for the Kerr effect and neutron Bragg diffraction is achieved without invoking a singular electronic structure for ceramic superconductors [9, 10, 13, 14, 15].

Use of an acentic site symmetry in the paramagnetic phase by a magnetic ion is not a requirement for polar magnetism to develop. Taking Hg1201 as an example, Cu ions are in centrosymmetric sites in the parent, paramagnetic state described by the P4/mmm1′ space group. However, in the pseudo-gap phase with the symmetry Cm′m′m′, inversion is replaced by anti-inversion, $\bar{1}'$. The physical mechanism for this is the development of polar quadrupoles acting as order parameters that break the inversion at Cu sites. Figure 1 depicts the motif of Dirac quadrupoles in Hg1201. A bootstrap scenario that is well-established in the context of axial magnetism in crystals might have its counterpart in the emergence of Dirac multipoles from ions that occupy centrosymmetric sites in the chemical structure, as with Cu ions Hg1201. We refer to the tension between the crystal-field potential and the exchange interaction. The crystal-field potential experienced by an ion opposes ordering of axial magnetic dipoles, with a singlet ground-state forecast for materials such as praseodymium. Ordering of axial dipoles is favoured by an exchange interaction, however. In such a situation, a magnetic dipole emerges from the singlet state when exchange interactions exceed a critical value of the energy interval occupied by crystal-field states [16, 17]. In the case of YBCO, by contrast, inversion symmetry at sites occupied by Cu ions does not exist even in the parent structure. Condensation of Dirac quadrupoles in YBCO breaks time-reversal symmetry together with some rotational symmetry, i.e., space inversion does not need to be broken at the magnetic phase transition for it is already structurally broken.

Magnetic charge in the pseudo-gap phase, with spatial order that does not break translation symmetry, is absent in all available theories [18, 19, 20]. Laughlin states that his Hartree-Fock computation of the cuprate phase diagram uses the most general set of Fermi liquid parameters allowed by symmetry [18]. However, bathos ensues in so far as orbital current *antiferromagnetism* – yet to be observed – is identified with the pseudo-gap phase while a ferro-type motif of magnetic charge, due to Dirac multipoles, is not present. Likewise results from a phenomenological ansatz, based on analogies to the approach to Mott localization, by Rice *et al*. [21]. The Mott insulator and, also, the loop-current insulator proposed by Varma [22] do not exist in the logical construct promoted by Laughlin [18]. It is beyond reasonable doubt that available neutron diffraction data are not consistent with the loop-current insulator [9, 10].

As a valued step toward establishing magnetic charge from Dirac multipoles one can enquire how it emerges in a Landau theory. In this respect, a crucial point is that the symmetry of the system becomes polar in an applied magnetic field, favouring a presence of Lifshitz-type invariants in Landau free-energy decomposition. Theses invariants are well known to result in a formation of inhomogeneous incommensurate electronic states [23]. At the present time, there is a solid evidence of strong impact of magnetic field on the incommensurate charge-density wave in hole-doped cuprates and formation of new modulated states in magnetic fields in excess of 10 T [24, 25]. This behaviour can be a direct consequence of the lack of the inversion symmetry, imposed by the magneto-electric nature of the order parameter in the pseudo-gap phase. A further development of the phenomenological approach will depend on the progress of the quantitative structure refinement of these new modulated electronic states with a precise determination of their symmetries.

In view of the newly established presence of magnetic charge in ceramic, high-$T_c$ superconductors, it is appropriate to investigate the corresponding atomic states and the structure of the neutron scattering amplitude. While revelations from cuprates inform much of the ensuing discussion, because of their place in recent history, the discussion has applications to any magnetic material for which symmetry allows the existence of magnetic charge. By way of a timely example, we discuss Bragg diffraction by a magnetically ordered iridate. To date, the magnetic structure adopted by $Sr_2IrO_4$ has been inferred by use of a scattering amplitude composed of magnetic dipoles alone, ignoring magnetic quadrupoles, for example, and all Dirac multipoles.

The magnetic polar interaction operator in neutron scattering is explored in the following section. (A full calculation of matrix elements for neutron scattering is quite technical and final results expressed with atomic multipoles are unfortunately complicated [4]. We steer clear of technical stuff, as far as it is reasonable to do so.) Its strength is related to the strength of electric dipole transitions, which makes it an excellent handle on charge transfer. Results for the neutron scattering amplitude that might be adequate for many materials are also listed in Section 2. Section 3 is given over to the construction of Dirac multipoles that exist in the pseudo-gap phase of the cuprate superconductor Hg1201, and Section 4 is devoted to selection rules for excitations. Diffraction from magnetically ordered $Sr_2IrO_4$ is discussed in Section 5, with particular attention to likely contributions from parity-even quadrupoles, anapoles and Dirac quadrupoles. A discussion and conclusions appear in Section 6.

## 2. Magnetic neutron scattering

The magnetic scattering amplitude for neutrons $\mathbf{Q}_\perp = \{[\mathbf{k} \times (\mathbf{Q} \times \mathbf{k})]/k^2\}$ in which $\mathbf{k}$ is the difference between the primary and secondary wavevectors. The intermediate operator, $\mathbf{Q}$, we separate in two contributions with exclusive use of axial and polar magnetic operators $\mathbf{Q} = \{\mathbf{Q}^{(+)} + \mathbf{Q}^{(-)}\}$. The first contribution $\mathbf{Q}^{(+)}$ is axial (parity-even and a pseudo-dipole), and Schwinger derived the result $\mathbf{Q}^{(+)} \approx \{f(k)(2\mathbf{S} + \mathbf{L})/2\}$ for a small k [26]. An atomic form-factor is included, with $f(0) = 1$ as usual. By contrast, magnetic operators in the second contribution $\mathbf{Q}^{(-)}$ are polar (parity-odd) and there is some advantage in an expression $\mathbf{Q}^{(-)} = (i\boldsymbol{\kappa} \times \mathbf{D})$ with a unit vector $\boldsymbol{\kappa} = \mathbf{k}/k$. The magnetic (time-odd) dipole $\mathbf{D}$ has no matrix elements different from zero if magnetic ions that contribute to it occupy sites that are centres of inversion symmetry. In the absence of a centre of symmetry a matrix element such as $\langle 3d|\mathbf{D}|4p\rangle$ for a 3d-transition ion can be different from zero. Ligand orbitals contribute to matrix elements, in addition to intra-ion orbitals, and their contribution is likely to be a significant factor in materials with strong hybridization, which is known to exist between Cu-3d and O-2p states within $CuO_4$ plaquettes in a cuprate [18, 20].

By way of an orientation to $\mathbf{Q}^{(-)}$ consider the contribution made by electron spins through the orbital-spin operator $[\exp(i\mathbf{k} \cdot \mathbf{R}) \mathbf{S}]$ in $\mathbf{Q}$, where $\mathbf{R} = \mathbf{n} R$ is the position of the electron. This operator yields a contribution $[ik R (\boldsymbol{\kappa} \cdot \mathbf{n}) \mathbf{S}]$ to $\mathbf{Q}^{(-)}$, at the second level in an expansion of the exponential in the small quantity $(kR)$. It is useful to re-write this manifestly parity-odd contribution using a standard identity for the product of Cartesian components of vector quantities,

$$(\boldsymbol{\kappa} \cdot \mathbf{n}) S_\alpha = \sum_\beta \kappa_\beta S_\alpha n_\beta = \sum_\beta \kappa_\beta \{(1/3)\delta_{\alpha\beta} \mathbf{S} \cdot \mathbf{n} + (1/2) \sum_\gamma \varepsilon_{\alpha\beta\gamma} (\mathbf{S} \times \mathbf{n})_\gamma$$

$$+ (1/2) [S_\alpha n_\beta + S_\beta n_\alpha - (2/3) \delta_{\alpha\beta} \mathbf{S} \cdot \mathbf{n}]\}.$$

The magnetic monopole $(\mathbf{S} \cdot \mathbf{n})$ will not contribute to $\mathbf{Q}_\perp$, because it is multiplied by $\boldsymbol{\kappa}$. This selection rule on the monopole is explicit in a general result for orbital-spin $\mathbf{D}(s)$ given in an Appendix. The dipole $\sum \kappa_\beta \varepsilon_{\alpha\beta\gamma} (\mathbf{s} \times \mathbf{n})_\gamma = [\boldsymbol{\kappa} \times (\mathbf{S} \times \mathbf{n})]_\alpha$ yields $\mathbf{D}(s) \propto (\mathbf{S} \times \mathbf{n})$ as the orbital-spin contribution to dipole approximation to $\mathbf{D}$. Our definition of the quadrupole in the final contribution is standard and the trace is zero, and it is re-written in an even more useful form at a later stage. Note that the anapole and quadrupole contributions are the same order in the small quantity $(kR)$.

Evidently, a matrix element of $\mathbf{D}(s)$ is proportional to the average value of the electric dipole moment $\langle R \rangle$. Cu 3d and 4p states participate in an intra-ion contribution to the dipole strength. In the case of YBCO, where Cu ions use sites that are not centres of inversion symmetry, the intra-ion transition is said to be Laporte-forbidden. An atomic code calculation for $Cu^{2+}$: $3d^8 - 4p^1$ yields the value $\langle 3d| R |4p\rangle \approx 0.53\ a_o$ with $a_o$ the Bohr radius. To the extent that radial wavefunctions in a highly ionized ion are hydrogenic in form, $\langle 3d| R |4p\rangle$ is proportional to $1/Z_c$ where $Z_c$ is the effective core charge seen by the jumping electron. And an explicit calculation yields $\langle 3d| R |4p\rangle = 1.3\ (a_o/Z_c)$. Usually, a stronger contribution to $\langle R \rangle$ is ligand to metal charge transfer, using an electron localized in a ligand orbital and the central Cu ion, because the dipole transition is generally Laporte-allowed. The low-energy electronic degrees of freedom are primarily $pd\sigma$ antibonding O p and Cu 3d $(x^2 - y^2)$ orbitals in the $CuO_2$ layer. The same anion and cation states determine the hybridization matrix that is proportional to the wavefunction overlap of Cu (3d) and O ($2p_x$, $2p_y$) holes. (Denoting the amplitude of the Cu-O hybridization by t the antiferromagnetic exchange parameter is proportional to $t^4$ [27]). In the vicinity of a Cu ion the oxygen $2p\sigma$ wavefunction can be expanded in the basis of Cu 4p orbitals, resulting in $|2p_\sigma\rangle \to \rho |4p\rangle$ and $\rho^2$ as large as 0.40.

An absolute value of the dipole matrix element can be obtained from measured line strengths of transitions or calculations of electronic structure. The standard formula for $\langle R \rangle$ in terms of the dimensionless oscillator strength f is,

$$\{\langle R \rangle/a_o\}^2 = (3\nu_1/\nu_2) (R_\infty/E_o) f.$$

In this formula, two levels are separated in energy by $E_o$, $\nu_1$ is the degeneracy of the lower level, $\nu_2$ is the degeneracy of the upper level, and $R_\infty = 13.61$ eV. Ellipsometric spectra for semiconducting, antiferromagnetic $YBa_2Cu_3O_6$ shows a sharp feature at $E_o = 1.77$ eV with $f = 0.17 \pm 0.03$ that is assigned to charge transfer transitions from oxygen p-states to copper d-states in the Cu-O plane [28]. Both initial and final states were interpreted as admixtures of Cu and O. $E_o$ decreases and the width of the feature increases with increasing temperature, while f is independent of temperature, to a good approximation. The 1.77 eV feature is removed by the introduction of carriers, and it is absent in spectra gathered on a metallic, superconducting $YBa_2Cu_3O_{6.85}$ sample [29]. For the non-metallic sample the formulae yields an estimate $\{\langle R\rangle/a_o\}^2 \sim 1.1$ on using $(3\nu_1/\nu_2) = 1$. Intra-ion, or Laporte-forbidden, transitions in magnetic crystals occur with $f \sim 10^{-4}$, while charge transfer transitions occur with $f \sim 0.1$ as already mentioned for semiconducting, antiferromagnetic $YBa_2Cu_3O_6$ [28].

A complete evaluation of the contribution to $\mathbf{Q}^{(-)}$ from the orbital-spin operator is summarized in an Appendix. Contributions considered thus far actually comprise the first term in an expansion of a radial integral $(h_1) = \langle n_a l_a | j_1(kR) | n_b l_b \rangle$ where $j_1(kR)$ is a spherical Bessel function of order 1. Estimates of $(h_1)$ for isolated Cu and U ions have been reported [4].

To complete the dipole approximation for $\mathbf{D}$ we add to $(\mathbf{S} \times \mathbf{n})$ contributions from the pure orbital operator $[\exp(i\mathbf{k} \cdot \mathbf{R}) (\boldsymbol{\kappa} \times \mathbf{p})/k]$, where $\mathbf{p}$ is the momentum operator conjugate to $\mathbf{R}$. Matrix elements of this operator are technically more difficult to evaluate than those of the orbital-spin operator, because $\mathbf{p}$ operates on orbitals and, also, one must handle operators that do not commute, as in the orbital anapole $\boldsymbol{\Omega} = (\mathbf{L} \times \mathbf{n} - \mathbf{n} \times \mathbf{L})$. There are two purely orbital contributions to the dipole approximation for $\mathbf{D}$ with individual radial integrals. They have in common unbounded behaviour as k tends to zero. One finds [4],

$$\mathbf{D} \approx (1/2) [ i(g_1) \mathbf{n} + 3(h_1) \mathbf{S} \times \mathbf{n} - (j_0) \boldsymbol{\Omega}]. \qquad (2.1)$$

Operators $\mathbf{n}$, $\mathbf{S} \times \mathbf{n}$, and $\boldsymbol{\Omega}$ in (2.1) are Hermitian. Estimates of radial integrals $(g_1)$, $(h_1)$, $(j_0)$ for $3d^8$ - $4p^1$ have been reported [4], together with the standard approximation for the atomic form factor in the parity-even contribution to the neutron-electron interaction $f(k) = \langle 3d|j_0(kR)|3d\rangle$ [30].

Equation (A5) is a complete answer for $\mathbf{D}(s)$ in terms of magneto-electric operators created with $\mathbf{S}$ and $\mathbf{n}$ and denoted $H^{K'}_{Q'}$, with projections $Q'$ in the range $-K' \leq Q' \leq K'$. As it turned out, the successful analysis of neutron diffraction data for the high-$T_c$ materials Hg1201 and YBCO required anapole and quadrupole contributions to $\mathbf{D}(s)$ [6, 7]. In terms of $\mathbf{Q}^{(-)}$ the appropriate results are,

$$Q^{(-)}_x \approx (3i/2) (h_1) [\boldsymbol{\kappa} \times (\mathbf{S} \times \mathbf{n})]_x - (3i/\sqrt{5}) [\kappa_z B^2_1 - \kappa_x A^2_2 + i\kappa_y B^2_2],$$

$$Q^{(-)}_y \approx (3i/2) (h_1) [\boldsymbol{\kappa} \times (\mathbf{S} \times \mathbf{n})]_y - (3i/\sqrt{5}) [- \kappa_z A^2_1 + \kappa_y A^2_2 + i\kappa_x B^2_2], \qquad (2.2)$$

$$Q^{(-)}_z \approx (3i/2) (h_1) [\boldsymbol{\kappa} \times (\mathbf{S} \times \mathbf{n})]_z + (3i/\sqrt{5}) [\sqrt{(3/2)} \kappa_z A^2_0 + i\kappa_y A^2_1 - \kappa_x B^2_1].$$

Neglect of Dirac octupoles, etc., in (2.2) makes them approximations. We adopt a notation $H^{K'}_{\pm Q'} = A^{K'}_{Q'} \pm B^{K'}_{Q'}$, and explicit expressions for $H^2_{\pm Q'}$ are listed in an Appendix. Results (2.2) are used in Section 5 to discuss diffraction by a magnetic iridate.

## 3. Atomic states

The amplitude for neutron Bragg diffraction is a sum of multipoles that are expectation values of tensor operators, e.g., dipoles $\langle(2\mathbf{S} + \mathbf{L})\rangle$ and $\langle\mathbf{\Omega}\rangle$ where angular brackets $\langle…\rangle$ denote the time-average of the enclosed operator. These ground-state entities are constrained by symmetry operations in the point group of the site used by a magnetic ion (Neumann's Principle) [31]. Magneto-electric point groups include the operation $\bar{1}'$ that forbids the formation of axial multipoles: the point groups are $\bar{1}'$, 2/m', 2'/m, mmm', m'm'm', 4/m', 4'/m', 4/m'm'm', 4/m'mm, 4'/m'm'm, $\bar{3}'$, $\bar{3}'$m', $\bar{3}'$m, 6/m', 6'/m, 6/m'm'm', 6/m'mm, 6'/mmm', m'$\bar{3}'$, m'$\bar{3}'$m and m'$\bar{3}'$m'. The magnetic monopole is forbidden by a mirror operation, m, so it is absent among multipoles in nine of the 21 magneto-electric point groups.

With magnetism in the pseudo-gap phase of Hg1201 in mind, we continue the current discussion of atomic states by considering a calculation that illustrates restrictions on multipoles imposed by the point group m'm'm' set in the basis {(1, −1, 0), (1, 1, 0), (0, 0, 1)} with respect to the tetragonal parent. The basis is henceforth labelled (x, y, z), and m'm'm' possesses diad axes of rotation symmetry $2_x$, $2_y$, $2_z$, in addition to $\bar{1}'$ (magnetic point-groups are available at reference [32]). To be concrete, we will couch the argument in terms of tensor operators $H^K_Q$ introduced in the previous section of magnetic neutron scattering, but it applies to any Dirac multipole. The multipole $\langle H^K_Q \rangle$ is unchanged by $2_z$ if the projection $Q$ is an even integer, while it is unchanged by $2_x$ and $2_y$ if $\langle H^K_Q \rangle = (-1)^K \langle H^K_{-Q} \rangle = (-1)^K \langle H^K_Q \rangle^*$ [31]. In consequence, $\langle H^K_0 \rangle = 0$ for $K$ odd, and, in particular, the anapole $\langle(\mathbf{S} \times \mathbf{n})\rangle = 0$.

A Cu 3d orbital is a $(x^2 - y^2)$-type in crystal axes and (xy)-type with respect to the basis required by magnetic symmetry. One can take $|xy\rangle = i(|d, +2\rangle - |d, -2\rangle)$ with $|d, m\rangle$ a d-state of projection m. A spin state $|\sigma\rangle$ has $\sigma = 1/2$. For the purpose of illustration, the Cu 3d-state is taken to be a product $|xy\rangle|\sigma\rangle$, which is best expressed in coupled states $|d; J, M\rangle$ with $J = 3/2$ and 5/2, e.g., $|d, +2\rangle|\sigma\rangle = |d; 5/2, 5/2\rangle$. A non-zero value for $\langle H^K_Q \rangle$ is attributed to the presence of 4p-states mixed with 3d-states. (Recall that 4p-states may include ligand O 2p-states via a charge transfer mechanism.) 4p-states included in the groundstate based on $|xy\rangle|\sigma\rangle$ must be consistent with the symmetry-imposed restriction $Q$ even, which leads to consideration of $|p; j, -3/2\rangle$ and $|p; j, 1/2\rangle$ with $j = 3/2$. As we will shortly see, limiting 4p-states to $j = 1/2$ and $|p; j, 1/2\rangle$ is not consistent with the interpretation of neutron diffraction data for Hg1201, because the interpretation calls for Dirac quadrupoles $\langle H^2_0 \rangle$ and $\langle H^2_{+2} \rangle$ different from zero [7]. For a minimal ground-state we use,

$$|g\rangle = N_o [|xy\rangle|\sigma\rangle + i\alpha |p; j, -3/2\rangle + i\beta |p; j, 1/2\rangle]. \qquad (3.1)$$

The two mixing parameters, α & β, must be purely real to achieve the symmetry-imposed conditions $\langle H^K_0 \rangle = \langle g|H^K_0|g\rangle = 0$ for $K$ odd, and, also, $\langle H^K_Q \rangle = (-1)^K \langle H^K_{-Q} \rangle$. Normalization of $|g\rangle$ is satisfied by $N_o = [2 + \alpha^2 + \beta^2]^{-1/2}$. We go on to find,

$$\langle H^2_0 \rangle = \surd(2/3)\, N_o^2\, \alpha\, (h_1),\ \text{and}\ \langle H^2_{+2} \rangle = -\surd(1/3)\, N_o^2\, \beta\, (h_1). \qquad (3.2)$$

Thus, the Dirac quadrupole $\langle H^2_0 \rangle$ is proportional to the mixing parameter α, while $\langle H^2_{+2} \rangle$ is proportional to the mixing parameter β in the minimal ground-state. The quoted numerical factors in $\langle H^2_0 \rangle$ and $\langle H^2_{+2} \rangle$ arise from reduced matrix-elements [4]. Properties of Dirac quadrupoles, including their dependence on temperature, can be inferred from Stone's model [33]. Returning to the 4p-states in $|g\rangle$,

$$|p; 3/2, -3/2\rangle = |p, -1\rangle\, |-\sigma\rangle,$$

$$|p; 3/2, 1/2\rangle = \surd(2/3)\, |p, 0\rangle\, |\sigma\rangle + \surd(1/3)\, |p, +1\rangle\, |-\sigma\rangle. \qquad (3.3)$$

Occurrence of opposite spin states can be attributed to spin-orbit mixing within 4p-states.

## 4. Inelastic scattering

Excitation from a state containing magnetic charge can be achieved with a polar magnetic operator. To illustrate constraints imposed by symmetry on an inelastic scattering event we continue with m′m′m′ for the point group in the magnetic ground-state, and a magneto-electric operator $H^K_Q$. Results have potential application to experiments using inelastic neutron scattering, and resonant inelastic x-ray scattering (RIXS). In a RIXS experiment, $H^K_Q$ can represent the magnetic operator in the channel enhanced by an atomic electric dipole - electric quadrupole event (E1-E2) that has $K$ = 1, 2, 3. (Selection rules for RIXS using and E1-E1 event and a fixed 90° scattering angle are reviewed by Ament *et al.* [34]. The next generation of instrument will have a variable scattering angle, which will be a bonus in disentangling electronic processes in an energy profile that is typically overcrowded [35].)

A judicious choice of variables for a character table is,

$$\alpha^K_Q = \{H^K_Q + (-1)^Q H^K_{-Q}\},\ \beta^K_Q = (1/i)\{H^K_Q - (-1)^Q H^K_{-Q}\}.$$

Symmetry operations in m′m′m′ are mentioned in the previous Section, and they are listed in the top row of Table 1, namely, diads $2_x$, $2_y$, $2_z$, in addition to anti-inversion $\bar{1}'$ and anti-mirror operations $m_x′$, $m_y′$, $m_z′$. Essential results needed to construct the character table are,

$$2_x\, \alpha^K_Q = (-1)^{K+Q}\, \alpha^K_Q,\ 2_y\, \alpha^K_Q = (-1)^K\, \alpha^K_Q,\ 2_z\, \alpha^K_Q = (-1)^Q\, \alpha^K_Q,$$

$$2_x\, \beta^K_Q = -(-1)^{K+Q}\, \beta^K_Q,\ 2_y\, \beta^K_Q = -(-1)^K\, \beta^K_Q,\ 2_z\, \beta^K_Q = (-1)^Q\, \beta^K_Q.$$

From Table 1 we find, symmetry-allowed transitions include m′m′m′ → m′m′m′ via operators $\alpha^2_0$, $\alpha^2_2$ or $\beta^3_2$. For all remaining operators considered selection rules are symmetry-selective, with m′m′m′ → $2_\eta/m_\eta′$ and Cartesian label η = x, y or z.

## 5. Diffraction by an iridate

By way of illustrating expressions in (2.2) for the neutron scattering amplitude we apply them to diffraction by a magnetic iridate, which has been the subject of many experimental and theoretical studies. Interest in 5d-based iridates stems in part from similarities with the $j_{eff}$ = 1/2 model encouraged by a large spin-orbit coupling [36, 37]. Results from experiments using neutron Bragg diffraction have made a significant contribution to our present knowledge about the magnetic properties of $Sr_2IrO_4$ [38, 39]. However, all the results have been inferred from the simplest expression for the scattering amplitude without justification. The shortcomings are demonstrated here by appealing to a model [41] that gives a valid interpretation of data gathered with resonant x-ray Bragg diffraction, while the original interpretation offered by the authors has no merit [40].

The model in question is based on a chemical structure $I4_1/acd$ with Ir ions at sites (8a) that have site symmetry $\bar{4}$, which does not possess inversion symmetry. There is evidence to suggest that the chemical structure has a lower symmetry, and we will comment on how the suspected departure from $I4_1/acd$ influences our results for the scattering amplitude. The magnetic structure inferred thus far from experiments is depicted in Fig. 2. It is described by the magnetic space-group $P_Icca$ that possesses an anti-body-centre condition, with ordering wavevector (1, 1, 1), and an extinction rule $(h + k + l)$ odd applied to integer Miller indices $h$, $k$, $l$ [38, 41, 42]. Ir site symmetry $2'_c$ confines magnetic dipoles to the a-b plane, with dipole motifs +, −, +, − (a-component) and +, +, −, − (b-component).

Scattering amplitudes are derived directly from the result (2) in reference [41] for the electronic structure factor for Ir ions in the space-group $P_Icca$. Let us consider Bragg reflections $(0, \kappa_y, \kappa_z)$ where Cartesian coordinates (x, y, z) match cell edges a, b & c, with $\kappa_y \propto k$, $\kappa_z \propto l$. Retaining only dipoles and quadrupoles in the scattering amplitude yields $\langle \mathbf{Q}_\perp \rangle^{(+)} = (\langle Q_{\perp, x} \rangle^{(+)}, 0, 0)$ with,

$$\langle Q_{\perp, x} \rangle^{(+)} \approx - (-1)^{n+m} \sqrt{3} \, [(\sqrt{3}/2) \langle T^1_x \rangle + (\kappa_z^2 - \kappa_y^2) \langle T^2_{+1} \rangle''], \qquad (5.1)$$

where $k = (2m + 1)$, $l = (4n + 2)$. The Bragg spot (0, 1, 2) is strong because of the relatively small magnitude of the scattering wavevector, k, and a favourable orientation of the magnetic dipole [38, 39]. The axial dipole $\langle \mathbf{T}^1 \rangle$ is proportional to the magnetic moment, $\langle \boldsymbol{\mu} \rangle$, in the forward direction of scattering. More generally, the so-called dipole-approximation is,

$$\langle \mathbf{T}^1 \rangle \approx (1/3) \, [2 \langle \mathbf{S} \rangle \langle j_0(k) \rangle + \langle \mathbf{L} \rangle \{ \langle j_0(k) \rangle + \langle j_2(k) \rangle \}], \qquad (5.2)$$

where $\langle j_0(k) \rangle = \langle 5d | j_0(kR) | 5d \rangle$ and $\langle j_2(k) \rangle$ are standard radial integrals [30], with $f(k) \approx \langle j_0(k) \rangle$, $\langle j_0(0) \rangle = 1$ and $\langle j_2(0) \rangle = 0$. The quadrupole in (5.1) is proportional to $\langle j_2(k) \rangle$. In fact, an approximation using $\langle T^1_x \rangle \approx [f(k) \langle \mu_x \rangle / 3]$ and the assumption $\langle T^2_{+1} \rangle'' = \text{Im}.\langle T^2_{+1} \rangle = 0$ is the basis on which all experiments have been interpreted. Dirac multipoles that we discuss later have also been neglected to date.

Mention of multipoles with an even rank in magnetic Bragg diffraction might seem strange at first, because magnetic multipoles often have an odd rank. This is the case if the ground-state of the magnetic ion is specified by a total angular momentum, j. Multipoles with an even rank measure the mixing of states with different angular momenta. In this capacity, the existence of quadrupoles ($K = 2$) and hexadecapoles ($K = 4$) quantify the limitation of the $j_{eff} = 1/2$ model of iridates, for the model is valid in the absence of both j-mixing and concomitant tetragonal distortion of the IrO$_6$ complex [41]. Estimates of $\langle T^K_Q \rangle$ with $K$ even and $Q$ odd, as required by 2'$_c$, derived from a Ir-wavefunction that reproduces the measured saturation magnetic moment and resonant x-ray diffraction pattern [41], shows that they are not small at intermediate wavevectors where they make a significant contribution to the intensity of a Bragg spot [42]. A wavefunction of the same type has recently been exploited to simulate second-harmonic generation [43], although the theoretical analysis flounders on an antiquated knowledge of neutron diffraction [44].

Dirac multipoles contribute to (0, $\kappa_y$, $\kappa_z$) Bragg spots with Miller indices $k = (2m + 1)$, $l = 4n$. From (2) we find $\langle \mathbf{Q}_\perp \rangle^{(-)} = (\langle Q_{\perp, x} \rangle^{(-)}, 0, 0)$ with,

$$\langle Q_{\perp, x} \rangle^{(-)} \approx -(-1)^{n+m} \kappa_z \sqrt{(3/2)} [\langle H^1_y \rangle - \sqrt{(6/5)} \langle H^2_{+1} \rangle']. \qquad (5.3)$$

From (A4) we have $\langle H^1_y \rangle \propto \langle \mathbf{S} \times \mathbf{n} \rangle_y$ and $\langle H^2_{+1} \rangle' = \mathrm{Re}.\langle H^2_{+1} \rangle \propto \langle S_x n_z + S_z n_x \rangle$.

Amplitudes $\langle \mathbf{Q}_\perp \rangle^{(+)}$ and $\langle \mathbf{Q}_\perp \rangle^{(-)}$ for a general reflection ($\kappa_x$, $\kappa_y$, $\kappa_z$) are readily derived from the structure factor (2) in reference [41], but we report no more results. Expressions (5.1) and (5.3) suffice to illustrate that neutron diffraction likely contains more in the way of useful information on Sr$_2$IrO$_4$ than has been generally recognized.

A few nuclear Bragg spots observed in the paramagnetic phase of Sr$_2$IrO$_4$ are not indexed on I4$_1$/acd [39]. This finding is taken to imply a chemical structure with less symmetry, and specifically the loss of spatial correlation between Ir ions in the a-b plane with sites (8a) in I4$_1$/acd split into two groups of inequivalent sites. The correlation between ions in the a-b plane is represented by a mirror symmetry normal to the b-axis, and in expressions (2.2) for scattering amplitudes it demands $A^2_1 = 0$ and $B^2_1 \propto \langle H^2_{+1} \rangle'$. Removing the mirror symmetry results in $A^2_1 \propto \langle H^2_{+1} \rangle''$, while $\langle T^2_{+1} \rangle'$ is added to the parity-even amplitude $\langle \mathbf{Q}_\perp \rangle^{(+)}$. Multipoles in $\langle \mathbf{Q}_\perp \rangle^{(+)}$ and $\langle \mathbf{Q}_\perp \rangle^{(-)}$ with higher ranks − expressions (5.1) and (5.3) are limited to $K = 1$ and $K = 2$ − will likewise proliferate on removal of the mirror symmetry, although projections $Q$ remain restricted by 2'$_c$ to odd values.

## 6. Discussion and conclusions

Schwinger limited his analysis of magnetic neutron scattering to a very small wavevector, k, where magnetic effects in scattering can be most pronounced. Dirac multipoles formed from electric dipole and spin operators are encountered at the next level of approximation in an expansion in terms of k of the scattering amplitude, cf., Section 2. Whence, Dirac multipoles of this type, enabled by entangled orbital and spin degrees of freedom, are not included in Schwinger's result. Nor are they included in the next milestone in the development of a theory of magnetic neutron scattering, which is a paper by George Trammell published in 1953 [46]. In this calculation, Dirac multipoles of all types are excluded by the restriction to unpaired electrons in an atomic shell, whereas Dirac multipoles are parity-odd entities and unpaired electrons must belong to different atomic shells, as we illustrate in Section 3.

Specifically, Dirac multipoles are both time-odd and parity-odd (magnetic and polar), and likely the best known are a magnetic monopole and an anapole studied extensively with resonant x-ray Bragg diffraction [5, 8, 45, 47]. Properties of the orbital-spin polar interaction operator for neutron scattering are reported. It can be written in the form $\mathbf{Q}^{(-)} = [i\boldsymbol{\kappa} \times \mathbf{D}(s)]$, where $\boldsymbol{\kappa}$ is a unit vector in the direction of the scattering wavevector [4]. A general expression for $\mathbf{D}(s)$ is complemented by approximate values of $\mathbf{Q}^{(-)}$; approximate in the sense that the values omit operators in $\mathbf{D}(s)$ other than Dirac dipoles (anapoles) and quadrupoles, with quadrupoles depicted in Figure 1.

Magneto-electric operators in $\mathbf{D}(s)$ include radial integrals that depend on the magnitude of the scattering wave-vector, k. Radial integrals in $\mathbf{D}(s)$ are similar to the familiar atomic form factor in the standard dipole-approximation for scattering by axial dipole moments, $\mathbf{Q}^{(+)} \approx \{(1/2) f(k) \langle\boldsymbol{\mu}\rangle\}$, where $\langle\boldsymbol{\mu}\rangle$ is the dipole moment and the form factor f(k) is normalized at unity for k = 0 [26]. For a fixed k, radial integrals diminish in magnitude with increasing multipole rank, which leads to the reasonable expectation that Dirac dipoles and quadrupoles furnish an adequate representation of $\mathbf{Q}^{(-)}$ in many materials.

Explicit expressions for Dirac quadrupoles in $\mathbf{Q}^{(-)}$ are listed. Expectation values required for neutron Bragg diffraction are investigated in Section 3 using an atomic orbital that represents a possible configuration of electrons at Cu sites in the cuprate Hg1201.

The strength of scattering by anapoles and magneto-electric quadrupoles is related to charge transfer mechanisms, and intra-ion processes, that contribute to the matrix element of the electric dipole operator. A published calculation demonstrates that, bonding O ions in the Cu-O plane of cuprates propagate superexchange, which creates magnetism and superconductivity [18]. Magnitudes of magneto-electric multipoles have not been derived from experimental data gathered using neutron Bragg diffraction, although the feasibility is firmly established by the seminal experiments on Hg1201 and YBCO [9, 10]. A possible mechanism for the formation of magneto-electric quadrupoles in Hg1201 has been explored in a simulation of electronic structure [48].

Excitations away from a state of magnetic charge are subject to legislation derived from symmetry, albeit local or collective excitations. By way of an orientation to the impact and nature of selection rules on magnetic scattering, we report a comprehensive investigation of Hg1201. Results have value for future experiments using inelastic neutron scattering or resonant inelastic x-ray scattering (RIXS). Selection rules are symmetry-selective, with allowed final states uniquely labelled by their magnetic and spatial character.

To broaden the scope away from ceramic superconductors we examine magnetic Bragg diffraction by an iridate. Magnetically ordered $Sr_2IrO_4$ has been the subject of many investigations, including resonant x-ray Bragg diffraction [40], neutron diffraction [39], and second-harmonic generation [43]. We reiterate that neutron diffraction data might contain important information on the validity of the $j_{eff} = 1/2$ model [36, 37, 42] and disclose the fact that anapoles and Dirac quadrupoles may contribute to selected Bragg spots.

**Acknowledgements.** We benefitted from correspondence and discussions with Dr Urs Staub at the PSI (CH), and written comments by Professor T M McQueen (The John Hopkins University, USA). At the Diamond Light Source, Ltd (DLS) we had discussions with Dr Mirian García-Fernández, Dr Andrew Walters, and Dr Kejin Zhou who build a next-generation RIXS beam-line (I21). Calculation of the electric dipole was made by Professor Gerrit van der Laan (DLS). Useful advice on DFT calculations was received from Dr Nikitas Gidopoulos and Dr Stewart Clark (Durham University), and Dr Bernard Delley (PSI).

**Appendix. Orbital-spin neutron interaction**

The primary aim is to define the dipole **D**(s) introduced in Section 2. In the process, we introduce quantities and definitions that occur also in a general discussion of the orbital-spin and orbital contributions to the neutron-electron interaction [4]. In so doing, it is hoped to dovetail the current and previous work, which does not contain an explicit expression for **D**(s).

A compact definition of the spin-orbital multipole operator, $H^{K'}$, makes use of a tensor product defined as,

$$\{A^a \otimes B^b\}^K_Q = \sum_{\alpha,\beta} A^a{}_\alpha B^b{}_\beta \, (a\alpha b\beta | KQ), \tag{A1}$$

where $A^a$ and $B^b$ are arbitrary spherical tensors. The Clebsch-Gordan coefficient and Wigner 3-j symbol in (A1) are related by [49 - 52],

$$(a\alpha b\beta | KQ) = (-1)^{-a+b-Q} \sqrt{(2K+1)} \begin{pmatrix} a & b & K \\ \alpha & \beta & -Q \end{pmatrix}. \tag{A2}$$

Two tensor products created from dipoles $a = b = 1$ are $\{A^1 \otimes B^1\}^0 = -(1/\sqrt{3})\,(\mathbf{A} \cdot \mathbf{B})$ for $K = Q = 0$, and $\{A^1 \otimes B^1\}^1 = (i/\sqrt{2})\,(\mathbf{A} \times \mathbf{B})$ for $K = 1$. In the present case, the two commuting tensors are spin, **S**, and a spherical harmonic, $C^K(\mathbf{n})$, normalized such that $C^1(\mathbf{n}) = \mathbf{n}$ the electric dipole operator. A matrix element $\langle n_a l_a | C^K(\mathbf{n}) | n_b l_b \rangle$ is zero unless $(l_a + l_b + K)$ is even, so $K$ is an odd integer for a parity-odd process. A triangle rule imposes $|l_a - l_b| \leq K \leq (l_a + l_b)$. We employ [4],

$$H^{K'}(K) = (-i)^{1+K+K'} \sqrt{[(2K+1)(2K'+1)/3]}\,(h_K)\,\{\mathbf{S} \otimes C^K(\mathbf{n})\}^{K'}, \tag{A3}$$

where the phase factor makes $H^{K'}(K)$ an Hermitian operator. The radial integral in (A3) is $(h_K) = \langle n_a l_a | j_K(kR) | n_b l_b \rangle$, with $(h_1) \approx (k/3)\,\langle n_a l_a | R | n_b l_b \rangle$ for small k, and notation makes explicit the dependence of the multipole operator on $K$. Of immediate interest for discussion in the main text are,

$$\mathbf{H}^1(1) = -\sqrt{(3/2)}\,(h_1)\,(\mathbf{S} \times \mathbf{n}), \quad H^2{}_{Q'}(1) = \sqrt{5}\,(h_1)\,\{\mathbf{S} \otimes \mathbf{n}\}^2{}_{Q'},$$

and,

$$\{\mathbf{S} \otimes \mathbf{n}\}^2{}_{\pm 2} = (1/2)\,[S_x n_x - S_y n_y \pm i(S_x n_y + S_y n_x)],$$

$$\{\mathbf{S} \otimes \mathbf{n}\}^2{}_{\pm 1} = - (1/2) [\pm (S_x n_z + S_z n_x) + i(S_y n_z + S_z n_y)],$$

$$\{\mathbf{S} \otimes \mathbf{n}\}^2{}_0 = (1/\sqrt{6}) [2S_z n_z - S_y n_y - S_x n_x]. \qquad (A4)$$

One can then show that,

$$\mathbf{D}(\mathbf{s}) = \sum_{K, K', x} (-1)^{1+K} i^{1+K'} (2x+1) \sqrt{[6(2K+1)]}$$

$$\times \begin{pmatrix} 1 & K & x \\ 0 & 0 & 0 \end{pmatrix} \begin{Bmatrix} K & K' & 1 \\ 1 & 1 & x \end{Bmatrix} \{\mathbf{H}^{K'}(K) \otimes \mathbf{C}^x(\boldsymbol{\kappa})\}^1. \qquad (A5)$$

A derivation of (A5) makes use of re-coupling of angular momenta, and key identities are found in references [50] and [52]. The 3-j symbol is zero unless $(K + x)$ is odd, and the 6-j symbol embodies the triangle rule $K' = K - 1, K, K + 1$. Note that $K' = 0$ is forbidden, whence the magnetic monopole $(\mathbf{S} \cdot \mathbf{n})$ is forbidden in neutron scattering, although it contributes to resonant x-ray scattering [8]. For $K$ odd $x = 0, 2$ for $K' = 1$; $x = 2$ for $K' = 2$; etc. Results in Section 2 are proportional to $(h_1)$ and can be derived from (A5) using $K = 1$ and $K' = 1, 2$.

It should be noted that $\mathbf{D}$ is arbitrary to within a function proportional to $\boldsymbol{\kappa} = \mathbf{k}/k$, because such a function does not contribute in $\mathbf{Q}_\perp = [i\boldsymbol{\kappa} \times \mathbf{D}]$. By way of an example, the term with $K = K' = 1$ in (A5) includes a contribution proportional to $(\boldsymbol{\kappa} [\boldsymbol{\kappa} \cdot \mathbf{H}^1])$ that does not survive in $\mathbf{Q}_\perp$.

**References.**

**Table 1.** Entries in the table tell if an operation in the point group m′m′m′, shown in the top row, leaves operators $\alpha^K_Q$ or $\beta^K_Q$ unchanged (entry = 1) or changes the sign of the operator (entry = −1). Operations from m′m′m′ that leave $\alpha^K_Q$ or $\beta^K_Q$ unchanged determine the symmetry of a state in an allowed transition using $\alpha^K_Q$ or $\beta^K_Q$. E.g., initial and final states connected by $\alpha^2_1$ must have symmetries m′m′m′ and $2_y/m_y$′.

| $\alpha^K_Q/\beta^K_Q$ | 1 | $2_x$ | $2_y$ | $2_z$ | -1′ | $m_x$′ | $m_y$′ | $m_z$′ | Sel. Rul. |
|---|---|---|---|---|---|---|---|---|---|
| $\alpha^1_1$ | 1 | 1 | -1 | -1 | 1 | 1 | -1 | -1 | $2_x/m_x$′ |
| $\beta^1_1$ | 1 | -1 | 1 | -1 | 1 | -1 | 1 | -1 | $2_y/m_y$′ |
|  |  |  |  |  |  |  |  |  |  |
| $\alpha^2_0$ | 1 | 1 | 1 | 1 | 1 | 1 | 1 | 1 | $m_x$′$m_y$′$m_z$′ |
| $\alpha^2_1$ | 1 | -1 | 1 | -1 | 1 | -1 | 1 | -1 | $2_y/m_y$′ |
| $\beta^2_1$ | 1 | 1 | -1 | -1 | 1 | 1 | -1 | -1 | $2_x/m_x$′ |
| $\alpha^2_2$ | 1 | 1 | 1 | 1 | 1 | 1 | 1 | 1 | $m_x'm_y'm_z'$ |
| $\beta^2_2$ | 1 | -1 | -1 | 1 | 1 | -1 | -1 | 1 | $2_z/m_z'$ |
| | | | | | | | | | |
| $\alpha^3_0$ | 1 | -1 | -1 | 1 | 1 | -1 | -1 | 1 | $2_z/m_z'$ |
| $\alpha^3_1$ | 1 | 1 | -1 | -1 | 1 | 1 | -1 | -1 | $2_x/m_x'$ |
| $\beta^3_1$ | 1 | -1 | 1 | -1 | 1 | -1 | 1 | -1 | $2_y/m_y'$ |
| $\alpha^3_2$ | 1 | -1 | -1 | 1 | 1 | -1 | -1 | 1 | $2_z/m_z'$ |
| $\beta^3_2$ | 1 | 1 | 1 | 1 | 1 | 1 | 1 | 1 | $m_x'm_y'm_z'$ |
| $\alpha^3_3$ | 1 | 1 | -1 | -1 | 1 | 1 | -1 | -1 | $2_x/m_x'$ |
| $\beta^3_3$ | 1 | -1 | 1 | -1 | 1 | -1 | 1 | -1 | $2_y/m_y'$ |

**Fig. 1.** Ferro-type ordering of Dirac quadrupoles in the Cu-O plane for Hg1201 in the pseudo-gap phase. Arrows indicate spin directions in magnetic, polar quadrupoles $\langle H^2_0 \rangle \propto \langle 3S_z n_z - \mathbf{S} \cdot \mathbf{n} \rangle$ and $\langle H^2_{+2} \rangle' \propto \langle S_x n_x - S_y n_y \rangle$ together with their response to spatial or time inversion. Oxygen ions are not shown; the four O ions in the Cu-O plane lie along the a- and b-axes at positions ($\pm 1/2, 0, 1/2$) and ($0, \pm 1/2, 1/2$). Basis $\{(1, -1, 0), (1, 1, 0), (0, 0, 1)\}$ with respect to the tetragonal parent P4/mmm, labelled (x, y, z) in the main text, and lobes of $\langle H^2_{+2} \rangle'$ are orientated at 45° with respect to cell edges a & b.

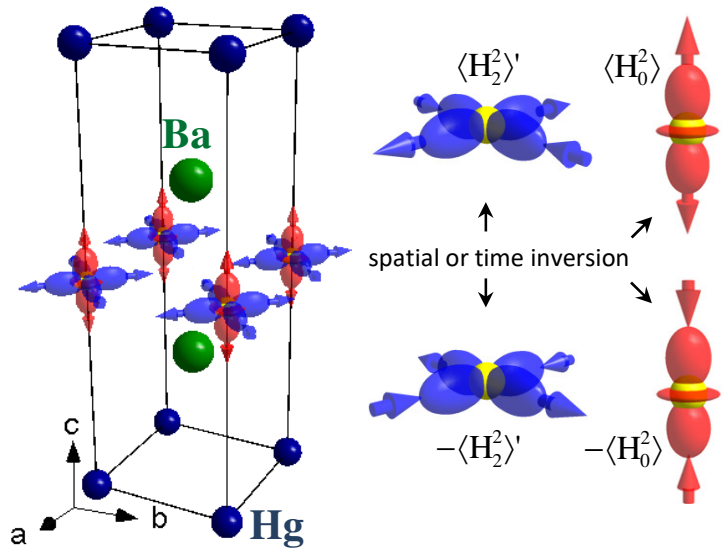

**Fig. 2**. Motif of magnetic dipoles in Sr$_2$IrO$_4$ that corresponds to a bi-dimensional irreducible representation M$_4$ in a magnetic space-group P$_I$cca, specified in terms of the Miller and Love notation [41, 42].

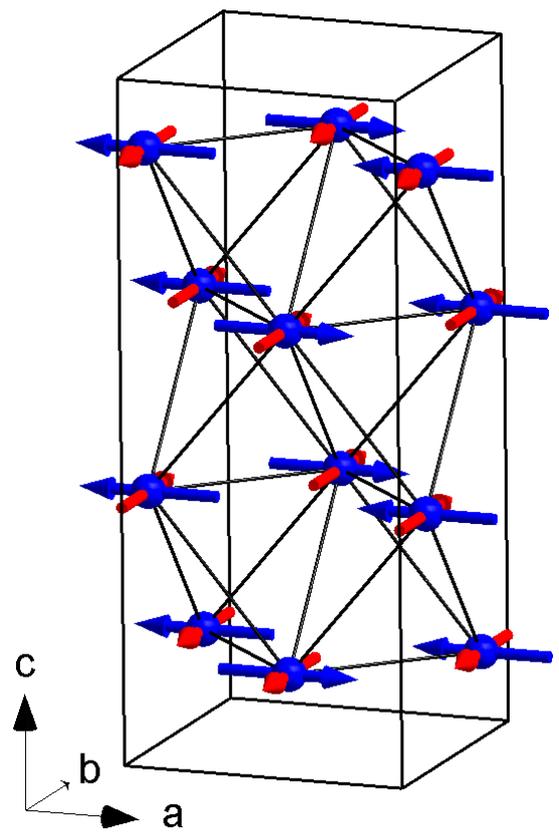